\documentclass[pra, showpacs, twocolumn, floatfix,superscriptaddress]{revtex4}
\usepackage{graphicx}
\usepackage{epstopdf}
\usepackage{amsmath, amsfonts, amssymb, bm}

\def\mpik{Max-Planck-Institut f\"ur Kernphysik, Saupfercheckweg 1, D-69117
Heidelberg, Germany}

\def\ifa{Institute of Applied Physics, Academy of Sciences of Moldova,
Academiei str. 5, MD-2028 Chi\c{s}in\u{a}u, Moldova}
\begin{document}
%%%%%%%%%%%%%%%%%%%%%%%%%%%%%%%%%%%%%%%%%%%%%%%%%%%%%%%%%%%%%%%%%
\title{Cavity quantum interferences with three-level atoms}
%%%%%%%%%%%%%%%%%%%%%%%%%%%%%%%%%%%%%%%%%%%%%%%%%%%%%%%%%%%%%%%%%
\author{Victor \surname{Ceban}}
\email{victor.ceban@phys.asm.md}
\affiliation{\mpik}
\affiliation{\ifa}

\author{Mihai A. \surname{Macovei}}
\email{macovei@phys.asm.md}
\affiliation{\mpik}
\affiliation{\ifa}
%%%%%%%%%%%%%%%%%%%%%%%%%%%%%%%%%%%%%%%%%%%%%%%%%%%%%%%%%%%%%%%%%
\date{\today}
%%%%%%%%%%%%%%%%%%%%%%%%%%%%%%%%%%%%%%%%%%%%%%%%%%%%%%%%%%%%%%%%%
\begin{abstract}
We discuss quantum interference phenomena in a system consisting from a laser driven three-level ladder-type emitter possessing orthogonal transition 
dipoles and embedded in a leaking optical resonator. The cavity mean-photon number vanishes due to the destructive nature of the interference 
phenomena. The effect occurs for some particular parameter regimes which were identified. Furthermore, upper bare-state population inversion occurs 
as well.
\end{abstract}
%%%%%%%%%%%%%%%%%%%%%%%%%%%%%%%%%%%%%%%%%%%%%%%%%%%%%%%%%%%%%%%%%%%
\pacs{42.25.Hz, 42.50.Ct, 42.50.Lc}
\maketitle
%%%%%%%%%%%%%%%%%%%%%%%%%%%%%%%%%%%%%%%%%%%%%%%%%%%%%%%%%%%%%%%%%%%

%%%%%%%%%%%%%%%%%%%%%%%%%%%%%%%%%%%%%%%%%%%%%%%%%%%%%%%%%%%%%%%%%%%
\section{Introduction}
%%%%%%%%%%%%%%%%%%%%%%%%%%%%%%%%%%%%%%%%%%%%%%%%%%%%%%%%%%%%%%%%%%%
Quantum interference effects involving various atomic transition amplitudes were intensively investigated recently \cite{agw,ficek,martin}.
It originate from indistinguishability of the corresponding transition pathways. As a consequence, quenching of spontaneous emission 
occurs due to quantum interference effects between decaying pathways which are dependent on mutual orientation of corresponding 
transition dipoles \cite{chk1}. Laser- or phase-control of spontaneous emission processes were demonstrated there. Also, quantum 
interference effects lead to very narrow spectral lines in the spontaneous emission spectrum of pumped molecular, semiconductor or 
highly charged ion systems \cite{chk2}. Furthermore, such spontaneously generated coherences in a large ensemble of nuclei operating 
in the x-ray regime and resonantly coupled to a common cavity environment were experimentally demonstrated in \cite{je}. 
Previously  it was shown that multi-level atoms interacting with the vacuum of a preselected cavity mode, in the bad-cavity limit, exhibit 
cavity induced quantum interference which is similar to the spontaneously generated coherences due to parallel transition dipoles 
\cite{knight}. Quantum interference effects in an ensemble of $^{229}{\rm Th}$ nuclei interacting with coherent light were demonstrated  
too, in Ref.~\cite{aaa}. Destructive or constructive interference effects were observed even in a strongly pumped few-level 
quantum- dot sample \cite{njp,viorel}. Protection of bipartite entanglement \cite{das} or continuous variable entanglement \cite{gx} via 
quantum interferences were shown to occur as well. Finally, electromagnetic induced transparency is an another phenomenon of quantum 
destructive interference which makes a resonant opaque medium highly transparent and dispersive within a narrow spectral band \cite{eit}. 

Based on this background, here, we study the quantum dynamics of a three-level ladder-type atomic system 
possessing orthogonal transition dipoles and embedded in an optical leaking resonator. The atomic sample is pumped coherently and 
resonantly with external electromagnetic field sources. Despite of photon scattering into surrounding electromagnetic modes including 
the cavity one and resonant laser-atom driving, we identify regimes when the cavity mode is empty in the good-cavity limit. We 
demonstrate that this occurs due to destructive quantum interference effects among the involved transition pathways. The effect is 
maximal when the cavity mode frequency is in resonance with particular resonance fluorescence sidebands. This is distinct from 
interference phenomena observed in \cite{je} where the experiment was performed in the bad-cavity limit and different parameter 
regimes. Furthermore, most upper bare-state population inversion is achieved as well in our system although it is identified with 
coherent population trapping effects rather than quantum interference phenomena which lead to zero cavity mean-photon numbers. 
The inversion can facilitate the generation of correlated or entangled photon-pairs when one photon lies in an optical range while 
another one is in a higher frequency domain, i.e., EUV or X ray etc.

The article is organized as follows. In Section II we describe the Hamiltonian, the adopted approximations and the master equation describing our 
system. In Section III we discuss and analyze the obtained results. The Summary is given in Section IV.
%%%%%%%%%%%%%%%%%%%%%%%%%%%%%%%%%%%%%%%%%%%%%%%%%%
\section{Theoretical framework}
\begin{figure}[t]
\centering
\includegraphics[width= 8cm ]{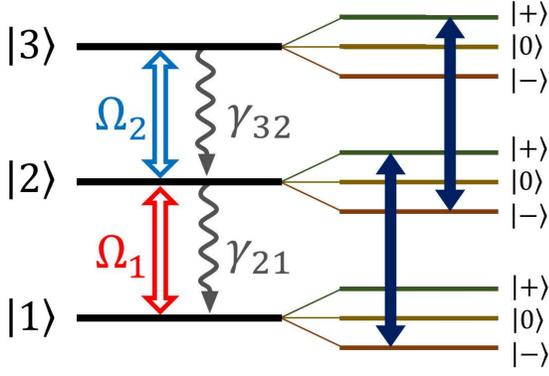}
\caption{(color online) Schematic of the model. A three-level atom, placed in a leaking optical cavity, interacts with external coherent fields with  
$\Omega_{2,1}$ being the corresponding Rabi frequencies. $\gamma_{32}$ and $\gamma_{21}$ are the respective spontaneous 
emission decay rates. The full arrows depict the dressed-state transitions in resonance with the cavity mode frequency  leading to cavity quantum 
interference phenomena.
\label{fig1} }
\end{figure}

%%%%%%%%%%%%%%%%%%%%%%%%%%%%%%%%%%%%%%%%%%%%%%%%%
We consider a laser driven three-level $\Sigma  \rm{-type}$ atom placed in a cavity of frequency $\omega_{c}$ and leaking rate $\kappa$. We 
denote via $\omega_{32}$ the transition frequency from the most excited level $ \vert 3 \rangle $ to the intermediate level $\vert 2\rangle$ 
whereas $\omega_{21}$ is the frequency of the transition from the state $\vert 2\rangle$ to the ground state $\vert 1\rangle$. The atom's 
decay rates via spontaneous emissions corresponding to each transition are defined as $\gamma_{32}$ and $\gamma_{21}$, respectively. 
The atomic system is coherently pumped, perpendicularly to the cavity axis, by two different laser fields of frequencies $\omega_{L1}$ and 
$\omega_{L2}$ with $\Omega_{1}$ and $\Omega_{2}$ being the corresponding Rabi frequencies on transitions 
$\vert 2 \rangle \leftrightarrow \vert 1 \rangle $ and $\vert 3 \rangle \leftrightarrow \vert 2 \rangle$ (see Fig.~\ref{fig1}). Correspondingly, 
the cavity-atom interaction constants are given by $g_{1}$ and $g_{2}$. The Hamiltonian describing the whole system is:
%%%%%%%%%%%%%%%%%%%%%%%%%%%%%%%%%%%%%%%%%%%%%%%%%%%%%%
\begin{eqnarray}
H &=& \hbar\omega_{c} a^{\dagger}a + \hbar \sum^{3}_{i=1} \omega_{i}S_{ii} + i \hbar g_{1} (a^{\dagger} S_{12} - S_{21} a ) \nonumber \\ 
&+& i\hbar g_{2} (a^{\dagger} S_{23} - S_{32} a ) + \hbar \Omega_{1}(S_{21}e^{-i \omega_{L1} t} + S_{12}e^{i \omega_{L1} t}) \nonumber \\ 
&+& \hbar \Omega_{2}(S_{32}e^{-i \omega_{L2} t} + S_{23}e^{i \omega_{L2} t}).
\label{Htot}
\end{eqnarray}
%%%%%%%%%%%%%%%%%%%%%%%%%%%%%%%%%%%%%%%%%%%%%%%%%%%%%%
Here $S_{ij} = \vert i \rangle \langle j \vert$, $\{i,j = 1,2,3 \}$, are the atom operators, while $\{ a^{\dagger},a\}$ are the cavity field creation 
and annihilation operators and obey the commutation relations $[S_{\alpha,\beta},S_{\beta', \alpha'}] = \delta_{\beta ,\beta'}S_{\alpha,\alpha'}
-\delta_{\alpha',\alpha}S_{\beta',\beta}$ and $[a, a^{\dagger}] = 1$, respectively. The first two terms in Eq.~(\ref{Htot}) describe the single 
mode free cavity field and the atom free Hamiltonians. The external laser fields and the cavity field interact with both transitions of the atom and 
every interaction is defined via separate terms corresponding to each transition. Thus, the next two terms of the Hamiltonian represent the 
interaction of the quantized cavity with the atom whereas the last two terms describe the laser-atom semi-classical interaction. 
%%%%%%%%%%%%%%%%%%%%%%%%%%%%%%%%%%%%%%%%%%%%%%%%%%%%%%%%%%%%%%
%%%%%%%%%%%%%%%%%%%%%%%%%%%%%%%%%%%%%%%%%%%%%%%%%%%%%%%%%%%%%%

The system's quantum dynamics is given by the master equation for the density operator $\rho$ as:
%%%%%%%%%%%%%%%%%%%%%%%%%%%%%%%%%%%%%%%%%%%%%%%%%%%%%%
\begin{eqnarray}
\frac{\partial \rho}{\partial t} &=& - \frac{i}{\hbar} [ H, \rho ] + \frac{\kappa}{2} \mathcal{L} (a) + \frac{\gamma_{32}}{2} 
\mathcal{L} (S_{23})+ \frac{\gamma_{21}}{2} \mathcal{L} (S_{12}), \nonumber \\
\label{Meq}
\end{eqnarray}
%%%%%%%%%%%%%%%%%%%%%%%%%%%%%%%%%%%%%%%%%%%%%%%%%%%%%%
where, on the right-hand side of the equation, the first term represents the coherent evolution based on the system Hamiltonian $H$, while the other 
terms describe the damping phenomena, i.e., the cavity field damping and the spontaneous emissions processes, respectively. The damping effects are 
expressed by the Liouville superoperator $ \mathcal{L} $, which acts on a given operator $\mathcal{O}$ as: 
$\mathcal{L} (\mathcal{O}) = 2 \mathcal{O} \rho \mathcal{O}^{\dagger} - \mathcal{O}^{\dagger} \mathcal{O} \rho - \rho \mathcal{O}^{\dagger} 
\mathcal{O}$.

In the usual interaction picture with $\omega_{L1}=\omega_{L2} \equiv \omega_{L}$ it is more convenient to apply the semi-classical dressed-state 
transformation defined as:
%%%%%%%%%%%%%%%%%%%%%%%%%%%%%%%%%%%%%%%%%%%%%%%%%%%%%%
\begin{eqnarray}
\vert 1 \rangle &=& -\frac{1}{\sqrt{2}} \cos{\theta}\, \vert -\rangle - \sin{\theta} \, \vert 0 \rangle + \frac{1}{\sqrt{2}} \cos{\theta} \, 
\vert + \rangle , \nonumber \\ 
\vert 2 \rangle &=&  \frac{1}{\sqrt{2}} \, \vert -\rangle  + \frac{1}{\sqrt{2}} \, \vert + \rangle ,  \nonumber \\ 
\vert 3 \rangle &=& -\frac{1}{\sqrt{2}} \sin{\theta} \, \vert -\rangle + \cos{\theta} \, \vert 0 \rangle + \frac{1}{\sqrt{2}} \sin{\theta}\, 
\vert + \rangle, 
\label{db}
\end{eqnarray}
%%%%%%%%%%%%%%%%%%%%%%%%%%%%%%%%%%%%%%%%%%%%%%%%%%%%%%
with $\theta = \arccos{(\Omega_{1}/\Omega)}$ and $\Omega=\sqrt{\Omega_{1}^{2}+\Omega_{2}^{2}}$. Then, in the dressed-state picture, 
the system's dynamics is described by the following dressed-state master equation:
%%%%%%%%%%%%%%%%%%%%%%%%%%%%%%%%%%%%%%%%%%%%%%%%%%%%%%
\begin{eqnarray}
\frac{\partial \rho}{\partial t} &=& - \frac{i}{\hbar} [ H, \rho ]+ \frac{\kappa}{2} \mathcal{L} (a) +  \frac{\gamma_{32} \cos^{2}{\theta}}{4}
(\mathcal{L} (R_{-0}) +\mathcal{L} (R_{+0}))  \nonumber \\
&+& \frac{\gamma_{32} \sin^{2}{\theta} +\gamma_{21}\cos^{2}{\theta}}{8} (\mathcal{L} (R_{z}) +\mathcal{L} (R_{+-})\nonumber \\ 
&+& \mathcal{L} (R_{-+})) + \frac{\gamma_{21} \sin^{2}{\theta}}{4} (\mathcal{L} (R_{0-}) +\mathcal{L} (R_{0+})) . \label{MEQdb}
\end{eqnarray}
%%%%%%%%%%%%%%%%%%%%%%%%%%%%%%%%%%%%%%%%%%%%%%%%%%%%%%
Here, the secular approximation was performed in the spontaneous emission terms by neglecting the time-dependent rapidly oscillating terms -
an approximation valid as long as $\gamma_{32(21)}/\Omega \ll 1$.  
The dressed-state atomic operators $R_{ij}=\vert i \rangle \langle j \vert$, $\{i,j\} \in \{-, 0, +\}$, and $R_{z} = R_{++} - R_{--}$, obey the 
same commutation relations as the old ones. The system Hamiltonian in the dressed-state picture is expressed as:
%%%%%%%%%%%%%%%%%%%%%%%%%%%%%%%%%%%%%%%%%%%%%%%%%%%%%%
\begin{eqnarray}
H &=& \hbar \delta_{c} a^{\dagger}a + \hbar \Omega R_{z} +  \lbrace  \frac{i}{2}(g_{1} \cos{\theta}+g_{2}\sin{\theta})a^{\dagger} R_{z}  
\nonumber \\ 
&+& \frac{i}{2} (g_{1} \cos{\theta} - g_{2} \sin{\theta}) a^{\dagger} (R_{+-}-R_{-+})\nonumber \\ 
&-& \frac{i}{\sqrt{2}} g_{1} \sin{\theta} \, a^{\dagger} (R_{0-}+R_{0+}) \nonumber \\
&-& \frac{i}{\sqrt{2}} g_{2} \cos{\theta} \, a^{\dagger} (R_{-0}+R_{+0})  \rbrace + \rm{H.c.} \rbrace, \label{Hdb}
\end{eqnarray}
%%%%%%%%%%%%%%%%%%%%%%%%%%%%%%%%%%%%%%%%%%%%%%%%%%%%%%
with $ \delta_{c} = \omega_{c} - \omega_{L}$.

As a next step, we perform a unitary transformation: $U(t)=\exp\{itH_{0}/\hbar\}$ where $H_{0}= \hbar \delta_{c} a^{\dagger}a + 
\hbar \Omega R_{z}$ to arrive at the following transformed Hamiltonian:
%%%%%%%%%%%%%%%%%%%%%%%%%%%%%%%%%%%%%%%%%%%%%%%%%%%%%%
\begin{eqnarray}
H&=& \frac{i}{2} (g_{1} \cos{\theta} - g_{2} \sin{\theta}) a^{\dagger}e^{i \delta_{c} t}  (R_{+-} e^{2i\Omega t} - R_{-+}e^{-2i\Omega t}) 
\nonumber \\ 
&-& \frac{i}{\sqrt{2}}  g_{1} \sin{\theta} \, a^{\dagger} e^{i \delta_{c} t}  (R_{0-}e^{i\Omega t} + R_{0+}e^{-i\Omega t}) 
\nonumber \\
&-& \frac{i}{\sqrt{2}} g_{2} \cos{\theta} \, a^{\dagger} e^{i \delta_{c} t} (R_{-0}e^{-i\Omega t} + R_{+0}e^{i\Omega t}) \nonumber \\
&+& \frac{i}{2}  (g_{1} \cos{\theta} + g_{2}\sin{\theta})a^{\dagger} e^{i \delta_{c} t} R_{z}  + \rm{H.c.}. \label{Hinter}
\end{eqnarray}
%%%%%%%%%%%%%%%%%%%%%%%%%%%%%%%%%%%%%%%%%%%%%%%%%%%%%% 
Analyzing the Hamiltonian one can conclude that the atom's resonance fluorescence spectra on each transition is formed of sidebands centered 
at $\omega_{L} \pm \Omega$ and $\omega_{L} \pm 2\Omega$ as well as a central peak around $\omega_{L}$. The dressed-state Hamiltonian 
significantly simplifies if one tunes the laser-cavity detuning $\delta_{c}$ in resonance with one of the sidebands frequency. In what follows, 
we shall consider such a situation when $\omega_{c}=\omega_{L} + 2\Omega$, i.e. $\delta_{c} = 2\Omega$ (notice that similar results would 
be obtained for a cavity tuned to the lowest energy sideband, i.e., when $\delta_{c} = -2\Omega$). In this case, the dressed-state Hamiltonian 
is
%%%%%%%%%%%%%%%%%%%%%%%%%%%%%%%%%%%%%%%%%%%%%%%%%%%%%%
\begin{eqnarray}
H = ig(a^{\dagger}R_{-+}  -  R_{+-}a) ,
\label{Hfinal}
\end{eqnarray}
%%%%%%%%%%%%%%%%%%%%%%%%%%%%%%%%%%%%%%%%%%%%%%%%%%%%%%
with $ g= \frac{1}{2} ( g_{2} \sin{\theta} - g_{1} \cos{\theta} )$. This Hamiltonian accurately describes the quantum dynamics within the 
adopted approximations as long as $g_{1,2}/\Omega \ll 1$. 
Furthermore, the obtained Hamiltonian has a similar form to the Hamiltonian of a two-level atom interacting with a 
cavity with an effective coupling $g$. This effective coupling originates from the quantum interference of the two dressed-state transition 
amplitudes, see Fig.~(\ref{fig1}), contributing to the atom pumping of the cavity mode. As it will be shown later, it is possible to configure 
the two Rabi frequencies in order to effectively decouple the atom from the cavity, i.e., the effective coupling vanishes when 
$g_{1}/g_{2} = \Omega_{2}/\Omega_{1}$.  Correspondingly, the cavity mode is empty, although the pumped atom spontaneously scatters 
photons. Notice that when $\delta_{c} = \{0, \pm \Omega\}$ there are no destructive quantum interference effects.

The master equation (\ref{MEQdb}), with the Hamiltonian (\ref{Hfinal}), is solved by projecting into the system's states basis \cite{hf}. The 
first projection in the atom's dressed-state basis leads to a set of five linearly coupled differential equations defined by the variables: 
$\rho^{(0)} = \rho_{--}+\rho_{00}+\rho_{++}$, 
$\rho^{(1)}= \rho_{++}+\rho_{--}$, 
$ \rho^{(2)}= \rho_{++}-\rho_{--}$, 
$\rho^{(3)} = (a^{\dagger} \rho_{+-}+\rho_{-+}a)/2$, and 
$\rho^{(4)} = (\rho_{+-} a^{\dagger} +a\rho_{-+})/2 $, where $\rho_{ij}=\langle i \vert \rho \vert j \rangle$, $\{i,j \in -, 0, +\}$, 
namely,
%%%%%%%%%%%%%%%%%%%%%%%%%%%%%%%%%%%%%%%%%%%%%%%%%%%%%%
\begin{eqnarray}
\dot{\rho}^{(0)}  &=& -2 g(\rho^{(4)}-\rho^{(3)}) + \frac{\kappa}{2} \mathcal{L}_{0}(a), \nonumber \\
\dot{\rho}^{(1)} &=& -2 g(\rho^{(4)}-\rho^{(3)}) + \frac{\kappa}{2} \mathcal{L}_{1}(a) -\frac{\alpha}{2} \rho^{(1)} 
+ \gamma_{32} \cos^{2}{\theta} \rho^{(0)} , \nonumber \\
\dot{\rho}^{(2)} &=& -2 g(\rho^{(4)}+\rho^{(3)}) + \frac{\kappa}{2} \mathcal{L}_{2}(a) - \frac{\beta}{2} \rho^{(2)}, 
\nonumber \\
\dot{\rho}^{(3)} &=& \frac{g}{4}(2 a^{\dagger} (\rho^{(2)}+\rho^{(1)}) a +a^{\dagger} a (\rho^{(2)} - \rho^{(1)}) + (\rho^{(2)} - \rho^{(1)})  
\nonumber \\ 
& \times& a^{\dagger} a ) - \zeta \rho^{(3)} + \frac{\kappa}{2}( \mathcal{L}_{3}(a) + \rho^{(3)} - 2 \rho^{(4)}) , \nonumber \\ 
\dot{\rho}^{(4)} &=& \frac{g}{4}(2 a (\rho^{(2)}-\rho^{(1)}) a^{\dagger} + a a^{\dagger} (\rho^{(2)}+\rho^{(1)}) + (\rho^{(2)}+\rho^{(1)})  
\nonumber \\
& \times & a a^{\dagger}) - \zeta \rho^{(4)} + \frac{\kappa}{2}( \mathcal{L}_{4}(a) - \rho^{(4)}). \label{SYSdb}
\end{eqnarray}
%%%%%%%%%%%%%%%%%%%%%%%%%%%%%%%%%%%%%%%%%%%%%%%%%%%%%%
Here $\alpha = \gamma_{21}\sin^{2}{\theta} + 2 \gamma_{32}\cos^{2}{\theta} $, 
$\beta = \gamma_{21} + \gamma_{32} \sin^{2}{\theta}$ and 
$\zeta = (\gamma_{21}(2+\cos^{2}{\theta}) + 3 \gamma_{32} \sin^{2}{\theta})/4$. 
The next projection in the field's basis leads to a set of infinite equations corresponding to the infinite Fock states 
$\lbrace\vert n\rangle, n\in \mathcal{N}\rbrace$, that is,
%%%%%%%%%%%%%%%%%%%%%%%%%%%%%%%%%%%%%%%%%%%%%%%%%%%%%%
\begin{eqnarray}
\dot{P}^{(0)}_{n}  &=& -2 g(P_{n}^{(4)}-P_{n}^{(3)}) + \kappa (n+1) P_{n+1}^{(0)} - \kappa n P_{n}^{(0)} , \nonumber \\
\dot{P}^{(1)}_{n} &=& -2 g(P_{n}^{(4)}-P_{n}^{(3)}) + \kappa (n+1) P_{n+1}^{(1)} -( \kappa n +\alpha/2) P_{n}^{(1)} \nonumber \\
&+& \gamma_{32} \cos^{2}{\theta} P_{n}^{(0)}, \nonumber \\
\dot{P}^{(2)}_{n} &=& -2 g(P_{n}^{(4)}+P_{n}^{(3)}) + \kappa (n+1) P_{n+1}^{(2)} - (\kappa n + \beta/2) P_{n}^{(2)} , \nonumber \\ 
\dot{P}^{(3)}_{n} &=& g n (P_{n-1}^{(1)} - P_{n}^{(1)} +P_{n-1}^{(2)} +P_{n}^{(2)})/2 - \kappa P_{n}^{(4)} \nonumber \\
&+& \kappa (n+1) P_{n+1}^{(3)} - (\kappa (n-1/2) + \zeta) P_{n}^{(3)}  , \nonumber \\ 
\dot{P}^{(4)}_{n} &=& g (n+1)(P_{n+1}^{(2)} + P_{n}^{(2)} - P_{n+1}^{(1)} + P_{n}^{(1)})/2  \nonumber \\
&+& \kappa (n+1) P_{n+1}^{(4)}- (\kappa (n+1/2) + \zeta) P_{n}^{(4)},  \label{SYSfin}
\end{eqnarray}
%%%%%%%%%%%%%%%%%%%%%%%%%%%%%%%%%%%%%%%%%%%%%%%%%%%%%%
where $P_{n}^{(i)}= \langle n\vert \rho^{(i)} \vert n\rangle $. Interestingly, $P_{n}^{(0)}$ are the diagonal elements of the field's reduced 
density matrix, i.e., it contains the trace over the atom's dressed-states: $P_{n}^{(0)} = \langle n \vert Tr_{atom}[\rho] \vert n \rangle$. 
Hence, it is possible to deduce the cavity field's mean photon number $\langle n \rangle$ by tracing over the the field's states:
%%%%%%%%%%%%%%%%%%%%%%%%%%%%%%%%%%%%%%%%%%%%%%%%%%%%%%
\begin{eqnarray}
\langle n \rangle = \langle a^{\dagger} a \rangle  =  \sum_{n=0}^{\infty} n P_{n}^{(0)}.
\label{n}
\end{eqnarray}
%%%%%%%%%%%%%%%%%%%%%%%%%%%%%%%%%%%%%%%%%%%%%%%%%%%%%%
Respectively, the second order photon-photon correlation function $g^{(2)}(0)$ is given by:
%%%%%%%%%%%%%%%%%%%%%%%%%%%%%%%%%%%%%%%%%%%%%%%%%%%%%%
\begin{eqnarray}
g^{(2)}(0) = \frac{\langle a^{\dagger} a^{\dagger} a a \rangle}{\langle n \rangle^{2}} 
=  \frac{\sum_{n=0}^{\infty} n (n-1) P_{n}^{(0)} }{\langle n \rangle^{2}}. \label{g2}
\end{eqnarray}
%%%%%%%%%%%%%%%%%%%%%%%%%%%%%%%%%%%%%%%%%%%%%%%%%%%%%%
After some mathematical manipulations one finds:
%%%%%%%%%%%%%%%%%%%%%%%%%%%%%%%%%%%%%%%%%%%%%%%%%%%%%%
\begin{eqnarray}
\langle R_{\pm \pm} \rangle =  \sum_{n=0}^{\infty} (P_{n}^{(1)} \pm P_{n}^{(2)})/2,
\label{sdb}
\end{eqnarray}
%%%%%%%%%%%%%%%%%%%%%%%%%%%%%%%%%%%%%%%%%%%%%%%%%%%%%%
with the condition $\langle R_{++} \rangle + \langle R_{00} \rangle + \langle R_{--}\rangle = 1$. Using also transformation 
(\ref{db}), the population of the upper bare state is given by:
%%%%%%%%%%%%%%%%%%%%%%%%%%%%%%%%%%%%%%%%%%%%%%%%%%%%%%
\begin{eqnarray}
\langle S_{33}\rangle = \cos^{2}{\theta} + (1 - 3 \cos^{2}{\theta}) \sum_{n=0}^{\infty}P_{n}^{(1)}/2.
\label{sbb}
\end{eqnarray}
%%%%%%%%%%%%%%%%%%%%%%%%%%%%%%%%%%%%%%%%%%%%%%%%%%%%%%

Finally, in order to solve the infinite system of Eqs.~(\ref{SYSfin}), we truncate it at a certain maximum value $n= N_{max}$ of 
considered Fock states.  The photon distribution of the field converges to zero for larger $N_{max}$, and thus, $ N_{max} $ is 
selected such that a further increase of its value does not modify the obtained results. 

%%%%%%%%%%%%%%%%%%%%%%%%%%%%%%%%%%%%%%%%%%%%%%%%%%
\section{Results and Discussion}
\begin{figure}[t]
\centering
\includegraphics[width= 8cm ]{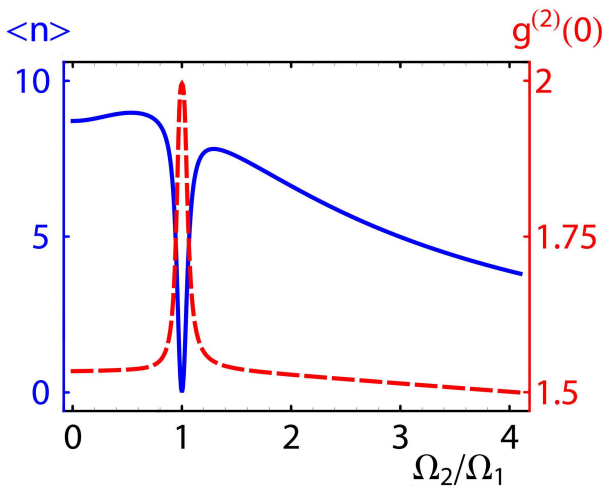}
\caption{\label{fig2} 
(color online) The quantum steady-state behaviors of the cavity mean photon number $ \langle n \rangle$ (solid curve) as well as 
of the second-order photon-photon correlation function $g^{(2)}(0)$ (dashed line) as a function of the ratio of the two 
Rabi frequencies $\Omega_{2}/\Omega_{1}$. Here, $\gamma_{32}=\gamma_{21} \equiv \gamma$, $g_{1}/\gamma=5.001$, 
$g_{2}/\gamma=5$ and $\kappa/\gamma=10^{-3}$.}
\end{figure}

%%%%%%%%%%%%%%%%%%%%%%%%%%%%%%%%%%%%%%%%%%%%%%%%%
The mean cavity photon number as well as their quantum statistics are shown in Fig.~(\ref{fig2}). A dip in the photon number is clearly 
visible when $g_{1}/g_{2}=\Omega_{2}/\Omega_{1}$. It is due to quantum interference effects with a destructive nature on the cavity 
field photons. The interference occurs because both dressed-state transitions of the atom are coupled to the cavity mode leading to 
indistinguishable photon emission (see also Fig.~\ref{fig1}). A destructive quantum interference phenomenon is observed when the 
cavity is tuned to one of the external sidebands, i.e., $\omega_{c} = \omega_{L} \pm 2 \Omega$. Respectively, the atom decouples 
from the cavity field in this particular case. Thus, the zero cavity photon detection in this scheme is directly related to quantum interference 
phenomena. Quantum switching devises are feasible applications here, because, the mean-photon number abruptly changes from zero to a 
particular value which depends on the atom-cavity coupling. The photon statistics shows super-Poissonian behaviors. Particularly, 
$g^{(2)}(0) \to 2$ and $\langle n \rangle \to 0$ when $\Omega_{2}/\Omega_{1}=g_{1}/g_{2}$. Here, the involved parameters or 
their ratios can be determined when the cavity mean- photon number vanishes. Additional applications may be related to quantum networks 
where tools to control the involved processes are highly required \cite{kimble}.
%%%%%%%%%%%%%%%%%%%%%%%%%%%%%%%%%%%%%%%%%%%%%%%%%%%%%%%%%%%%%%
%%%%%%%%%%%%%%%%%%%%%%%%%%%%%%%%%%%%%%%%%%%%%%%%%%%%%%%%%%%%%%

\begin{figure}[t]
\centering
\includegraphics[width= 8cm ]{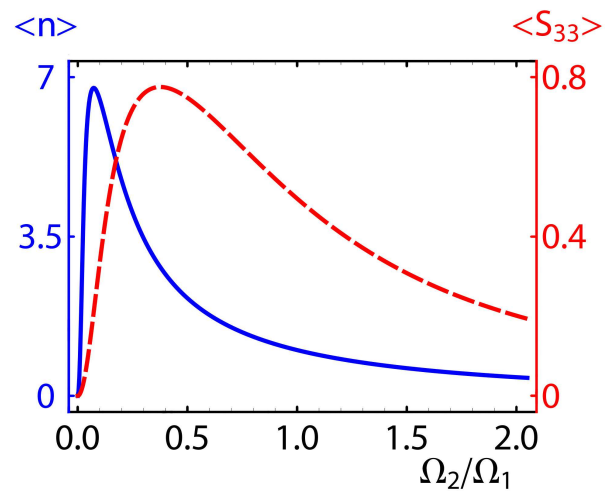}
\caption{\label{fig3} 
(color online) The steady-state evolution of the cavity mean photon number $\langle n \rangle$ (solid line) and the upper bare-state population  
$\langle S_{33}\rangle$ (dashed curve) as a function of $\Omega_{2}/\Omega_{1}$. Here, $\gamma_{32}/\gamma_{21}=10^{-2}$, 
$g_{2}/\gamma_{21}=5$, $\kappa/\gamma_{21}=10^{-3}$.}
\end{figure}

Fig.~\ref{fig3} shows the bare-state population of the upper state $|3\rangle$ as well as the cavity mean-photon number when 
$\gamma_{32} \ll \gamma_{21}$ while $\omega_{32}=\omega_{L2}$ and $\omega_{21}=\omega_{L1}$ with $\omega_{L1} 
\not=\omega_{L2}$. The cavity mode resonantly couples with the upper atomic transition $|3\rangle \leftrightarrow | 2\rangle$ 
only, i.e. $g_{1}=0$. This situation is described as well by the analytical formalism developed here via setting $g_{1}=0$ in 
Eqs.~(\ref{SYSfin}). Reasonable population inversion is achieved. We have found that the inversion is a signature of the coherent 
population trapping phenomenon \cite{tq}, i.e., the atom is trapped in the dressed-state $|0\rangle = \cos{\theta}|3\rangle - \sin{\theta}|1\rangle$. 
Then with a suitable chose of the ratio $\Omega_{2}/\Omega_{1}$ one can transfer the populations among the state 
$|3\rangle \leftrightarrow 1\rangle$. Furthermore, the cavity field does not affect the bare-state population dynamics in the adopted 
approximations. As a concrete atomic system, for this particular configuration, one may consider He atoms, 
$\{3^{1}S,2^{1}P_{1},1^{1}S_{0}\}$ \cite{scully}. The spontaneous decay rates ratio is approximately 
$\gamma_{32}/\gamma_{21} \approx 10^{-2}$. The corresponding transition wavelengths are $\lambda_{32}=728.3{\rm nm}$ and 
$\lambda_{21}=58.4{\rm nm}$. Instead of a continuous wave laser on the high-frequency transition one may consider a long pulse 
laser wave. Potential applications here may be related to entangled photon pair emissions \cite{sczb} in optical and EUV (or even higher)
frequency ranges. 

%%%%%%%%%%%%%%%%%%%%%%%%%%%%%%%%%%%%%%%%%%%%%%%%%%
\section{Summary}
%%%%%%%%%%%%%%%%%%%%%%%%%%%%%%%%%%%%%%%%%%%%%%%%%
Summarizing, we have investigated the quantum dynamics of a
three-level atom embedded in an optical cavity and resonantly
interacting with external coherent electromagnetic waves. We
have found parameter regimes where the atom completely decouples
from the interaction with the cavity field. As a consequence,
the cavity mean photon number goes to zero. Photon
vanishing is due to quantum interference effects involving two
possible dressed-state atomic transitions. Their indestinguishability
leads to destructive quantum interference phenomena.
Upper bare-state population inversion occurs as well.
%%%%%%%%%%%%%%%%%%%%%%%%%%%%%%%%%%%%%%%%%%%%%%%%%%%%%%%%%%%%%%
%%%%%%%%%%%%%%%%%%%%%%%%%%%%%%%%%%%%%%%%%%%%%%%%%%%%%%%%%%%%%%

%%%%%%%%%%%%%%%%%%%%%%%%%%%%%%%%%%%%%%%%%%%%%%%%%%%%%%%%%%%%%%%%%%%
%%%%%%%%%%%%%%%%%%%%%%%%%%%%%%%%%%%%%%%%%%%%%%%%%%
\section*{Acknowledgment}
%%%%%%%%%%%%%%%%%%%%%%%%%%%%%%%%%%%%%%%%%%%%%%%%%
We appreciate the helpful discussions with Professor Christoph H. Keitel and are grateful for the hospitality of the 
Theory Division of the Max Planck Institute for Nuclear Physics from Heidelberg, Germany.  
Furthermore, we acknowledge the financial support by the German Federal Ministry of Education and Research, grant No. 01DK13015, and 
Academy of Sciences of Moldova, grants No. 13.820.05.07/GF and 15.817.02.09F. 
%%%%%%%%%%%%%%%%%%%%%%%%%%%%%%%%%%%%%%%%%%%%%%%%%%%%%%%%%%%%%%%%%%

%%%%%%%%%%%%%%%%%%%%%%%%%%%%%%%%%%%%%%%%%%%%%%%%%%%%%%%%%%%%%%

%%%%%%%%%%%%%%%%%%%%%%%%%%%%%%%%%%%%%%%%%%%%%%%%%%%%%%%%%%%%%%

\begin{thebibliography}{33}
%%%%%%%%%%%%%%%%%%%%%%%%%%%%%%%%%%%%%%%%%%%%%%%%%%%%%%%%%%%%%%
\bibitem{agw} G. S. Agarwal, \rm{Quantum Statistical Theories of Spontaneous Emission and their Relation to other Approaches,} 
(Springer, Berlin, 1974).

\bibitem{ficek} Z. Ficek and S. Swain, \rm{Quantum Interference and Coherence: Theory and Experiments,} 
(Springer, Berlin, 2005).

\bibitem{martin} M. Kiffner, M. Macovei, J. Evers, and C. H. Keitel, \rm{Vacuum-Induced Processes in Multilevel Atoms,}
Progress in Optics {\bf 55}, 85 (2010).

\bibitem{chk1} S.-Y. Zhu and M. O. Scully, \rm{Spectral Line Elimination and Spontaneous Emission Cancellation via Quantum Interference,} 
Phys. Rev. Lett. {\bf 76}, 388 (1996);
M. A. G. Martinez, P. R. Herczfeld, C. Samuels, L.M. Narducci, and C. H. Keitel, \rm{Quantum interference effects in spontaneous atomic emission: 
Dependence of the resonance fluorescence spectrum on the phase of the driving field,} Phys. Rev. A {\bf 55}, 4483 (1997); 
E. Paspalakis and P. L. Knight, \rm{Phase Control of Spontaneous Emission,} Phys. Rev. Lett. {\bf 81}, 293 (1998).

\bibitem{chk2} C. H. Keitel, \rm{Narrowing Spontaneous Emission without Intensity Reduction,} Phys. Rev. Lett. {\bf 83}, 1307 (1999);
O. Postavaru, Z. Harman, and C. H. Keitel, \rm{High-Precision Metrology of Highly Charged Ions via Relativistic Resonance Fluorescence,} 
Phys. Rev. Lett. {\bf 106}, 033001 (2011).

\bibitem{je} K. P. Heeg, H.-C. Wille, K. Schlage, T. Guryeva, D. Schumacher, I. Uschmann, K. S. Schulze, B. Marx, T. K\"{a}mpfer, 
G. G. Paulus, R. R\"{o}hlsberger, and J. Evers, \rm{Vacuum-Assisted Generation and Control of Atomic Coherences at X-Ray Energies,} 
Phys. Rev. Lett. {\bf 111}, 073601 (2013).

\bibitem{knight} B. M. Garraway, and P. L.  Knight, \rm{Cavity modified quantum beats,} Phys. Rev. A {\bf 54}, 3592 (1996);
A. Patnaik, and G. S. Agarwal, \rm{Cavity-induced coherence effects in spontaneous emissions from preselection of polarization,} 
Phys. Rev. A {\bf 59}, 3015 (1999).

\bibitem{aaa} S. Das, A. P\'{a}lffy, and C. H. Keitel, \rm{Quantum interference effects in an ensemble of 
$^{229}{\rm Th}$ nuclei interacting with coherent light,} Phys. Rev. C {\bf 88}, 024601 (2013).

\bibitem{njp} B. D. Gerardot, D. Brunner, P. A. Dalgarno, K. Karrai, A. Badolato, P. M. Petroff, and R. J. Warburton, 
\rm{Dressed excitonic states and quantum interference in a three-level quantum dot ladder system,} 
New Journal of Physics {\bf 11}, 013028 (2009).

\bibitem{viorel} V. Ciornea, M. A. Macovei, \rm{Cavity-output-field control via interference effects,} Phys. Rev. A {\bf 90}, 043837 (2014).

\bibitem{das} S. Das, G. S. Agarwal, \rm{Protecting bipartite entanglement by quantum interferences,} Phys. Rev. A {\bf 81}, 052341 (2010).

\bibitem{gx} Z.-h. Tang, G.-x. Li, and Z. Ficek, \rm{Entanglement created by spontaneously generated coherence,} 
Phys. Rev. A {\bf  82}, 063837 (2010).

\bibitem{eit} S. E. Harris, \rm{Electromagnetically Induced Transparency,} Phys. Today {\bf 50}, 36 (1997); 
M. Fleischhauer, A. Imamoglu, and J. P. Marangos, 
\rm{Electromagnetically induced transparency: Optics in coherent media,} Rev. Mod. Phys. {\bf 77}, 633 (2005);
Y. Sun, Y. Yang, H. Chen, Sh. Zhu, \rm{Dephasing-Induced Control of Interference Nature in Three-Level Electromagnetically Induced 
Transparency System,} Sci. Rep. {\bf 5}, 16370 (2015).

\bibitem{hf} T. Quang, and H. Freedhoff, \rm{Atomic population inversion and enhancement of resonance fluorescence in a cavity,} 
Phys. Rev. A {\bf 47}, 2285 (1993).

\bibitem{kimble} H. J. Kimble, \rm{The quantum internet,} Nature (London) {\bf 453}, 1023 (2008).

\bibitem{tq} N. N. Bogolubov Jr., T. Quang, and A. S. Shumovsky, \rm{Double optical resonance in a system of atoms,} 
Phys. Lett. A {\bf 112}, 323 (1985).

\bibitem{scully} P. K. Jha, A. A. Svidzinsky, and M. O. Scully, \rm{Coherence enhanced transient lasing in XUV regime,} 
Laser Phys. Lett. {\bf 9}, 368 (2012).

\bibitem{sczb} M. O. Scully and M. S. Zubairy, \rm{Quantum Optics,} (Cambridge University Press, Cambridge, UK, 1997).
%%%%%%%%%%%%%%%%%%%%%%%%%%%%%%%%%%%%%%%%%%%%%%%%%%%%%%%%%%%%%%
\end{thebibliography}
\end{document}
%%%%%%%%%%%%%%%%%%%%%%%%%%%%%%%%%%%%%%%%%%%%%%%%%%%%%%%%%%%%%%